\documentclass[11pt]{article}

\usepackage{amscd}
\usepackage{amsmath}

\begin{document}

\newcommand{\N}{N\raise.7ex\hbox{\underline{$\circ $}}$\;$}

\title{The Lorentz Group, Noncommutative Space-Time,  and  Nonlinear
Electrodynamics in Majorana-Oppenheimer Formalism }
\author{V. Red'kov\footnote{redkov@dragon.bas-net.by}, E.Tolkachev\footnote{tea@dragon.bas-net.by}   \\[5mm]
Institute of Physics, National  Academy of Sciences of Belarus
 }

\date{}

\maketitle

\begin{abstract}

Non-linear electrodynamics arising in the frames of  field
theories in non-commuta\-ti\-ve space-time is examined on the base
of the Riemann-Silberstein-Majorana-Oppen\-heimer  formalism. The
problem of form-invariance of the non-linear constitutive
relations governed by six non-commutative  parameters $\theta_{kl}
\sim {\bf K} = {\bf n} + i {\bf m}$ is explored in detail on the
base of the complex orthogonal group theory SO(3.C). Two Abelian
2-parametric small groups, isomorphic to each other  in abstract
sense, and leaving unchangeable the extended constitutive
relations  at arbitrary six  parameters   $\theta_{kl}$ of
effective  media  have been found, their realization depends
explicitly on   invariant length   $ {\bf K}^{2}$. In the case of
non-vanishing length a special reference frame in which the small
group has the structure $SO(2) \otimes SO(1,1)$ has been found. In
isotropic case no such reference frame exists. The way to
interpret both  Abelian small groups in physical terms consists in
factorizing  corresponding Lorentz transformations into Euclidean
rotations and boosts. In the context  of general  study of various
dual symmetries  in non-commutative field theory,
  it is demonstrated explicitly  that
the non-linear constitutive equations in non-commutative
electro\-dynamics are not invariant under continuous dual
rotations, instead only invariance under  discrete dual
transformation exists.

\end{abstract}

\section{Introduction}

As known [1-15] interest  in field theory models in a
non-commutative space-time has been grown notably after creating
in \cite{Seiberg-Witten-1999}  a general algorithm to relate usual
Yang-Mills gauge models to their non-commutative counterparts.
There appears  a great deal of new physical problems to
investigate, besides the question of the hypothetic  coordinate
non-commutativity has become   of  practically  testable nature.
Noticeable progress in describing  symmetry of non-commutative
spaces  was achieved  on the base of twisted Poincare group.

For instance, the mapping by Seiberg -- Witten refers  the
non-commutative extension of electro\-dynamics  to the usual
microscopic Maxwell  theory with  special non-linear constitutive
relations. Examining all possible sym\-metries of these new
constitutive relations seems to be a  significant point in order
to discern  the effects of the space-time  non-commutativity in
observable electromagnetic    non-linear  effects.

The problem of form-invariance of the non-commutativity structural
equations (see below) was considered in the literature. Several
simple non-commutative parameters were listed which alow for
existence of some residual Lorentz symmetry -- the later is
recognized  to have the structure $SO(2) \otimes SO(1,1)$.

The aim of the present article is to establish  subgroups of the
Lorentz  group leaving form-invariant the commutator of space-time
coordinates with \underline{arbitrary} noncommutative
antisymmetric matrix. The starting commutative relationship
transform  with respect to Lorentz group according to
\begin{eqnarray}
[L_{k}^{\;\;\;a}x_{a},  L_{l}^{\;\;\;b}x_{b} ]_{-} = i\;
L_{k}^{\;\;\;a}  L_{l}^{\;\;\;b}  \theta_{ab} =i \theta'_{kl} \; .
\label{0}
\end{eqnarray}

There exist several different views   on the transforms of the
matrix $\theta^{\mu\nu}$. Evidently,  we aim at extension of
Lorentz invariant models in ordinary Minkowski space-time to
models in non-commutative space-time.

We might consider skew-symmetric object $\theta^{\mu\nu}$ just as
a tensor under the Lorentz group, without any physically
preferable reference frame. Therefore,  six parameters involved in
$\theta^{\mu\nu}$-entity depend on the choice of the reference
frame, they behave like all other tensor os spinor objects in
physics. Within that approach  any  field model in non-commutative
space-time must involve only Lorentz covariant   constructs.
Similar line of argument was used by Herman Minkowski when
creating microscopic electrodynamics in moving medium. As known,
according to Minkowski constitutive equations, Euclidean rotations
do not chance parameters of the uniform medium, $\epsilon$ and  $\mu$ ,
whereas all boost transform them into new ones depending on the
velocity vector $\vec{V}$ of the reference frame. Differently, it
sounds as follows: small Lorentz  group leaving  invariant
parameters of an uniform media coincides with  real orthogonal
group $SO(3.R)$. Below we consider a similar problem in the frames
of a non-commutative electrodynamics.

The most radical   attitude to the transforms of
$\theta^{\mu\nu}$-entity may be formulated as follows: six
parameters involved in $\theta^{\mu\nu}$-entity provide us with
new six fundamental constants. However,  im\-mediately one
questions may be posed: in with reference frame me must take these
fundamental constants. And then  what are symmetry transformations
(small Lorentz group) leaving invariant these six parameters. In a
sense, in this point we turn back to the old question on existence
of a fundamental ether. No solution for ether problem has found
till now, so it is hardly reasonable to  reanimate the old
unsolved  puzzle in new embodiment.

Evidently, presented below simple mathematical treatment is of
value in any case, irrespective of the choice between two
mentioned  views. As mentioned, several particular examples of
such small  (or stability)  subgroups    were noticed  in the
literature, so our analysis extends and completes  previous
considerations. In a sense, the problem  may be  straightforwardly
solved  with the help of old and well elaborated technique in the
theory of the Lorentz group \cite{Fedorov-1980},
\cite{Berezin-Kurochkin-Tolkachev-1989}. A basic tool used in this
article is the theory of complex rotation group SO(3.C),
isomorphic to the Lorentz group, and the theory of  the  special
linear group SL(2.C), spinor covering for Lorentz group. So to
deal with the non-linear Maxwell theory we employ the known
Riemann-Silberstein-Majorana-Oppenheimer  approach -- for more
detail  and references see \cite{Bogush-Red'kov-Tokarevskaya-Spix-2008}.

In the context  of general  study of various dual symmetries  in
non-commutative field theory one other
problem will be considered:  it is demonstrated explicitly  that
the known non-linear constitutive equations arising from
non-commutative  electrodynamics in  the first order approximation
are not invariant under continuous dual rotations, instead only
invariance under  discrete dual transformation exists.

\section{ Basic facts in the Lorentz group, notation}

Let us recall basic facts in the theory of the Lorentz group and
related to it,
 focusing on its parametrization \cite{Fedorov-1980}, \cite{Berezin-Kurochkin-Tolkachev-1989}.
Let us start with the real rotation group $SO(3.R)$  and its
covering  $SU(2)$:
\begin{eqnarray}
 B(n) = n_{0} - i {\bf n} \; \vec{\sigma} =   \cos
{\alpha \over 2} - i \sin {\alpha \over 2} \; {\bf e} \;
\vec{\sigma}  \; , \qquad
 n_{0}^{2} + {\bf n}^{2} = 1 \; ,
\nonumber
\\
O (n)  =  I + 2 \; [ \; n_{0} \;{\bf n} ^{\times} + ( {\bf n}
^{\times})^{2} \; ]\;  , \qquad  ( {\bf n} ^{\times})_{il}=
-\epsilon _{ilj} \;n _{j}  \; , \nonumber
\\
O(n)  =\left | \begin{array}{lll}
 1 -2 (n_{2}^{2} + n_{3}^{2})   &   -2n_{0}n_{3} + 2n_{1}n_{2}    &   +2n_{0}n_{2} + 2n_{1}n_{3}  \\
 +2n_{0}n_{3} + 2n_{1}n_{2}     &  1 -2 (n_{3}^{2} + n_{1}^{2})   &   -2n_{0}k_{1} + 2n_{2}n_{3}   \\
 -2n_{0}n_{2} + 2n_{1}n_{3}     &   +2n_{0}n_{1} + 2n_{2}n_{3}    &  1 -2 (n_{1}^{2} + n_{2}^{2})
 \end{array} \right | .
 \label{1b}
 \end{eqnarray}

\noindent The composition rule in the unitary group is
\begin{eqnarray}
n''_{0} = n'_{0} n_{0} - {\bf n}' {\bf n} \; ,  \qquad {\bf n}'' =
n_{0}' {\bf n} +  n_{0} {\bf n} ' +   {\bf n} '  \times {\bf n} \;
; \label{2a}
\end{eqnarray}

\noindent transition to explicit parametrization of the rotation
group is achieved by the introduction of the Gibbs' 3-vector (for
more details see in \cite{Fedorov-1980}):
\begin{eqnarray}
{\bf c} = { {\bf n} \over  n_{0} } = \mbox{tg} \;{\alpha \over 2}
\; {\bf e} \; , \qquad {\bf c}'' = {{\bf c}' + {\bf c} + {\bf c}'
\times {\bf c} \over 1 - {\bf c}' {\bf c}} \; , \qquad O ({\bf c})
=  I + 2 \; { \; {\bf c} ^{\times} + ( {\bf c} ^{\times})^{2} \;
 \over 1 + {\bf c}^{2} } \nonumber
\\
 = {1 \over  1 + {\bf c}^{2}} \; \left |
\begin{array}{lll}
 1  +{\bf c}^{2} -2 (c_{2}^{2} + c_{3}^{2})   &   -2c_{3} + 2c_{1}c_{2}    &   +2c_{2} + 2c_{1}c_{3}  \\
 +2c_{3} + 2c_{1}c_{2}     &  1 +{\bf c}^{2}  -2 (c_{3}^{2} + c_{1}^{2})   &   -2c_{1} + 2c_{2}c_{3}   \\
 -2c_{2} + 2c_{1}c_{3}     &   +2c_{1} + 2c_{2}c_{3}    &  1 +{\bf c}^{2}  -2 (c_{1}^{2} + c_{2}^{2})
 \end{array} \right | \; .
\nonumber
 \end{eqnarray}

\noindent One should note the peculiarity: if   $n_{0}=0$  (when
$\alpha =\pi$), then
\begin{eqnarray}
B(n) =     - i  \; {\bf n} \; \vec{\sigma} \; , \qquad {\bf c}
=\infty \; {\bf e} \; , \qquad O (\infty \; {\bf e})  =  I + 2 \;
( {\bf e} ^{\times})^{2}  \; .
\nonumber
\end{eqnarray}

\noindent Rotation matrices  (\ref{1b}) cab be written differently
through  $(\alpha, {\bf e})$.
\begin{eqnarray}
O(\alpha , {\bf e}) =
 \left | \begin{array}{rrr}
1 -  F (e_{2}^{2} + e_{3}^{2})        & -\sin \alpha  \;e_{3}
+ F e_{1}e_{2}    &    \sin \alpha \; e_{2} +  F  e_{1}e_{3}  \\
  \sin \alpha \; e_{3} + F  e_{1} e_{2}     &  1 - F (e_{3}^{2} + e_{1}^{2})   &
  -\sin \alpha \; e_{1} + F  e_{2}e_{3}   \\
 -\sin \alpha \; e_{2} + F e_{1}e_{3}     &   \sin \alpha \; e_{1} +  F e_{2}e_{3}    &
 1  - F(e_{1}^{2} + e_{2}^{2})
 \end{array} \right | ,
\label{4a}
\end{eqnarray}

\noindent where $F =(1-\cos \alpha) $;  at  $\alpha = \pi$ it
reads
\begin{eqnarray}
O =
 \left | \begin{array}{rrr}
1 -  2 (e_{2}^{2} + e_{3}^{2})        &  + 2  e_{1}e_{2}    &    2  e_{1}e_{3}  \\
  + 2  e_{1} e_{2}     &  1 - 2 (e_{3}^{2} + e_{1}^{2})   &
  +2  e_{2}e_{3}   \\
  + 2 e_{1}e_{3}     &   2  e_{2}e_{3}    &
 1  - 2(e_{1}^{2} + e_{2}^{2})
 \end{array} \right | = I + 2 \; ( {\bf e}
^{\times})^{2}  \; .
\nonumber
\end{eqnarray}

Extension to the special linear group $SL(2.C)$, spinor covering
for the (proper orthochronous) Lorentz group
 $L_{+}^{\uparrow}$, is achieved by   formal change   $(n_{0}, -i{\bf n})$ to  any complex $(k_{0}, {\bf k})$:
 \begin{eqnarray}
 B (k_{0}, {\bf k} ) = k_{0}  +\;  k_{j} \;
\sigma_{j}
= (n_{0} + i m_{0}) + (-i n_{j} +  m_{j}) \sigma_{j}\; ,\;
\nonumber
\\
\mbox{det}\; B =  k_{0}^{2}  - {\bf k}^{2}
= n_{0}^{2} + {\bf
n}^{2}  - m_{0}^{2}
 - {\bf m}^{2}  + 2i ( n_{0}m_{0} + {\bf n} {\bf m})
= 1  \label{5a}
\end{eqnarray}

\noindent  with  the following composition rule
\begin{eqnarray}
k''_{0} = k'_{0} k_{0} - {\bf k}' {\bf k} \; ,  \qquad {\bf k}'' =
k_{0}' {\bf k} +  k_{0} {\bf k} ' +   i {\bf k} '  \times {\bf k}
\nonumber
\end{eqnarray}

\noindent which coincides with (\ref{2a}) when restricting to
subgroup SU(2).

 The complex orthogonal group $SO(3.C)$ may be defined as $2
\rightarrow 1$ mapping from  $SL(2.C)$,  its  elements are
\begin{eqnarray}
 O (k)  =  I + 2 \; [ \; k_{0} \;{\bf k}
^{\times} + ( {\bf k} ^{\times})^{2} \; ]
  \nonumber
 \\
 =
 \left | \begin{array}{lll}
 1 +2 (k_{2}^{2} + k_{3}^{2})   &   -2ik_{0}k_{3} - 2k_{1}k_{2}    &   +2ik_{0}k_{2} - 2k_{1}k_{3}  \\
 +2ik_{0}k_{3} - 2k_{1}k_{2}     &  1 +2 (k_{3}^{2} + k_{1}^{2})   &   -2ik_{0}k_{1} - 2k_{2}k_{3}   \\
 -2ik_{0}k_{2} - 2k_{1}k_{3}     &   +2ik_{0}k_{1} - 2k_{2}k_{3}    &  1 +2 (k_{1}^{2} + k_{2}^{2})
 \end{array} \right | \;.
 \label{5c}
 \end{eqnarray}

\noindent Euclidean rotations are specified by
\begin{eqnarray}
k_{0}= n_{0}\; , \qquad k_{j} = -i n_{j}\; , \qquad  n_{0} = \cos
{\alpha \over 2}\; , \qquad
 {\bf n} =\sin {\alpha \over 2} \; {\bf e} \; ;
 \nonumber
 \end{eqnarray}

\noindent note identities
\begin{eqnarray}
 [O(n)]^{*} =  O(n) \; , \qquad  [O(n)] ^{-1} =
O(\bar{n}) = [O(n)]^{tr} \;. \nonumber
\end{eqnarray}

 \noindent
 Lorentz boosts are specified by
\begin{eqnarray}
k_{0} = n_{0} \; , \qquad {\bf k} =  {\bf m} \; , \qquad n_{0}=
\mbox{ch}\;{\beta \over 2} \; , \qquad {\bf m} = \mbox{sh}\;{\beta
\over 2} \;{\bf e}   \; ,\hspace{10mm} \nonumber
\\
O  =  \left | \begin{array}{rrr} 1 -  (1- \mbox{ch} \beta )
(e_{2}^{2} + e_{3}^{2})        & - i\; \mbox{sh} \beta  \;e_{3}
+ G e_{1}e_{2}    &    i\;  \mbox{sh} \beta  \; e_{2} +  G  e_{1}e_{3}  \\
i\; \mbox{sh} \beta  \; e_{3} + (1-\mbox{ch} \beta )  e_{1} e_{2}
&  1 - G (e_{3}^{2} + e_{1}^{2})   &
  - i \; \mbox{sh} \beta  \; e_{1} + G  e_{2}n_{3}   \\
 -i \; \mbox{sh} \beta  \; e_{2} +  G  e_{1}e_{3}     &   i\; \mbox{sh} \beta  \; e_{1} +  G  e_{2}e_{3}    &
 1  -  G(e_{1}^{2} + e_{2}^{2})
 \end{array} \right |
\nonumber
\end{eqnarray}

\noindent where $G = (1- \mbox{ch}\;  \beta  )$; note identities
\begin{eqnarray}
[O(n_{0}, {\bf m}) ] ^{*} = [O(n_{0}, {\bf m}) ] ^{-1} =  O(n_{0},
-{\bf m}) =[O(n_{0}, {\bf m}) ]^{tr} \; . \nonumber
\end{eqnarray}

Let us write down  the real Lorentz transformation over 4-vectors:
\begin{eqnarray}
L = \left | \begin{array}{rrrr} k_{0}k^{*}_{0} & (-k^{*}_{0}k_{1}-
k_{0}k^{*}_{1})  &
 -k^{*}_{0}k_{2} -  k_{0}k^{*}_{2}  & -k^{*}_{0}k_{3} -  k_{0}k^{*}_{3}  \\
 -k^{*}_{0}k_{1}  -k_{0}k^{*}_{1}  &  k_{0} k^{*}_{0}  &
 -ik^{*}_{0} k_{3} +  i k_{0}k^{*}_{3}  & +ik^{*}_{0}k_{2} -  ik_{0}k^{*}_{2}  \\
- k^{*}_{0}k_{2}  -k_{0}k^{*}_{2}  &  ik^{*}_{0}k_{3} -i
k_{0}k^{*}_{3}
   &  k_{0}k^{*}_{0}  & -ik^{*}_{0}k_{1}+ i  k_{0}k^{*}_{1}  \\
-k^{*}_{0}k_{3}  - k_{0}k^{*}_{3}  & -ik^{*}_{0}k_{2} + i
k_{0}k^{*}_{2}  &  +ik^{*}_{0}k_{1} -  ik_{0}k^{*}_{1}  &
k_{0}k^{*}_{0}
\end{array}  \right |
\nonumber
\end{eqnarray}
\begin{eqnarray}
+ \left | \begin{array}{rrrr} D_{0}& i(+k_{2}k^{*}_{3} -
k_{3}k^{*}_{2}) &
 i(-k_{1}k^{*}_{3}+k_{3}k^{*}_{1})  & i(k_{1}k^{*}_{2} - k_{2}k^{*}_{1}) \\
-i(+k_{2}k^{*}_{3} -k_{3}k^{*}_{2}) & D_{1}  &
k_{1}k^{*}_{2}+k_{2}k^{*}_{1}
& k_{1}k^{*}_{3} +k_{3}k^{*}_{1}  \\
-i(-k_{1}k^{*}_{3} +k_{3}k^{*}_{1})& k_{1}k^{*}_{2}+k_{2}k^{*}_{1}
&
D_{2}  & k_{2}k^{*}_{3}+k_{3}k^{*}_{2}   \\
-i(+k_{1}k^{*}_{2}-k_{2}k^{*}_{1} ) &
+k_{1}k^{*}_{3}+k_{3}k^{*}_{1}  &  +k_{2}k^{*}_{3}
 +k_{3}k^{*}_{2} &  D_{3}
 \end{array} \right | \;
\nonumber
\end{eqnarray}

\noindent where
\begin{eqnarray}
D_{0}=  k_{j}k^{*}_{j} \; , \; D_{1} =
k_{1}k^{*}_{1}-k_{2}k^{*}_{2}-k_{3}k^{*}_{3}\;,
\nonumber
\\
D_{2} = k_{2}k^{*}_{2}  - k_{1}k^{*}_{1} -k_{3}k^{*}_{3} \; ,
\qquad
\; D_{3} = k_{3}k^{*}_{3} -k_{1}k^{*}_{1}-k_{2}k^{*}_{2}
\nonumber
\end{eqnarray}

\noindent or taking into account   $k_{a}=-i n_{a} +  m_{a}$:
\begin{eqnarray}
L = 2 \left | \begin{array}{cccc} (n_{0}^{2} + m_{0}^{2})/2  &
n_{1}m_{0} - n_{0}m_{1}  &  n_{2}m_{0} - n_{0}m_{2} &
n_{3}m_{0} - n_{0}m_{3}  \\
n_{1}m_{0} - n_{0}m_{1}  & (n_{0}^{2} + m_{0}^{2})/2 &
 -n_{0}n_{3} -  m_{0}m_{3}  &  n_{0}n_{2} +  m_{0}m_{2}  \\
n_{2}m_{0} - n_{0}m_{2} & n_{0}n_{3} +  m_{0}m_{3}   & (n_{0}^{2}
+ m_{0}^{2})/2
 & - n_{0}n_{1} - m_{0}m_{1}  \\
n_{3}m_{0} - n_{0}m_{3} & -n_{0}n_{2} - m_{0}m_{2}  &  2
n_{0}n_{1} +  m_{0}m_{1}  & (n_{0}^{2} + m_{0}^{2})/2
\end{array}  \right |
\nonumber
\end{eqnarray}
\begin{eqnarray}
+  2 \; \left | \begin{array}{rrrr}
D_{0}/2 &  n_{2} m_{3} - n_{3} m_{2} &  n_{3} m_{1} - n_{1} m_{3}   &  n_{1} m_{2} - n_{2} m_{1}  \\
-n_{2} m_{3} + n_{3} m_{2}  & D_{1}/2  & n_{1} n_{2}+ m_{1}m_{2} & n_{1} n_{3}+ m_{1}m_{3}  \\
-n_{3} m_{1} + n_{1} m_{3} & n_{1} n_{2}+ m_{1}m_{2} &
D_{2}/2 &   n_{2} n_{3}+ m_{2}m_{3}   \\
- n_{1} m_{2} + n_{2} m_{1}& n_{1} n_{3}+ m_{1}m_{3}  & n_{2}
n_{3}+ m_{2}m_{3} & D_{3}/2
 \end{array} \right |
\nonumber
\end{eqnarray}

\noindent where
\begin{eqnarray}
D_{0} = n_{1}^{2} + m_{1}^{2} + n_{2}^{2} + m_{2}^{2} + n_{3}^{2}
+ m_{3}^{2} \; ,
\nonumber
\\
D_{1} = n_{1}^{2} + m_{1}^{2} - n_{2}^{2} - m_{2}^{2} - n_{3}^{2}
- m_{3}^{2} \; ,
 \nonumber
\\
D_{2} = -n_{1}^{2} - m_{1}^{2} + n_{2}^{2} + m_{2}^{2} - n_{3}^{2}
- m_{3}^{2} \;,
\nonumber
\\
 D_{3} = -n_{1}^{2} - m_{1}^{2} - n_{2}^{2}
- m_{2}^{2} + n_{3}^{2} + m_{3}^{2} \; . \nonumber
\end{eqnarray}

\noindent Let us verify these formulas for Lorentz boosts:
$n_{0} = \mbox{ch}{\beta \over 2} \; , \; {\bf m} =
\mbox{sh}{\beta \over 2} \; {\bf e} \; , \; {\bf e}^{2} = 1 \;
;$ the matrix $L$ reads
\begin{eqnarray}
L =
 \left | \begin{array}{cccc}
\mbox{ch}\; \beta   &  - \mbox{sh}\; \beta \; e_{1}  &  - \mbox{sh}\; \beta \; e_{2} & -\mbox{sh}\; \beta \; e_{3}   \\
-\mbox{sh}\; \beta \;  e_{1}   &  1 +(\mbox{ch}\; \beta -1)
e_{1}^{2}   &
(\mbox{ch}\; \beta -1 ) e_{1}e_{2} & (\mbox{ch}\; \beta -1 ) e_{1}e_{3}  \\
-\mbox{sh}\; \beta\;  e_{2} & (\mbox{ch}\; \beta -1 )e_{1}e_{2} &
1 +(\mbox{ch}\; \beta -1) e_{2}^{2}    &
(\mbox{ch}\; \beta -1 )e_{2}e_{3}   \\
- \mbox{sh}\; \beta \;  e_{3} & (\mbox{ch}\; \beta -1 )e_{1}e_{3}
&
 (\mbox{ch}\; \beta -1 )e_{2}e_{3} & 1 +(\mbox{ch}\; \beta -1) e_{3}^{2}
 \end{array} \right |  ,
\label{11d}
\end{eqnarray}

\noindent which in terms of space-time transformation coincides
with the standard form
\begin{eqnarray}
t' =  \mbox{ch} \;\beta \; t  - \mbox{sh}\; \beta \; ({\bf e} {\bf
x})\; , \qquad {\bf x}' = -\mbox{sh}\; \beta \;{\bf e} \; t + [ \;
{\bf x} + (\mbox{ch}\; \beta -1)\; {\bf e} \;  ({\bf e}{\bf x}) \;
]  \; . \nonumber
\end{eqnarray}

\section{The problem of a small group in  $SO(3.C)$,\\ non-isotropic case
}

 Let return to eq.  (\ref{5a})
 and note that the whole set of element of $SL(2.C)$    can be divided into two subsets
 depending on ${\bf k}$ with vanishing or not length. In this Section we consider
 the non-isotropic case,  ${\bf k}^{2}  \neq 0$. Here, one may introduce a $(\gamma, {\bf \Delta})$-parametrization of
 that subset as follows:
 \begin{eqnarray}
 B(k) = \sin {\gamma \over 2 } - i \; \sin {\gamma \over 2} \; {\bf \Delta}\; ,
\qquad  \gamma = \alpha + i \beta  \; ,\nonumber
\\
{\bf \Delta} = {\bf N} + i {\bf M}  , \; {\bf \Delta}^{2} = ({\bf
N}^{2} - {\bf M}^{2}) + 2i {\bf N} {\bf M} = 1  \; . \label{13a}
\end{eqnarray}

\noindent Now, Euclidean rotation and Lorentzian boost are
specified  respectively  by conditions:
\begin{eqnarray}
\beta = 0\;  , \;\;  {\bf M}= 0 \;  , \qquad \mbox{and}\qquad
\alpha  = 0\;  , \;\; {\bf M}= 0 \; . \nonumber
\end{eqnarray}

\noindent One can express the  Lorentz matrix $L$ in terms of the
variable $(\gamma , {\bf \Delta})$
\begin{eqnarray}
\gamma = \alpha + i\;  \beta \; , \qquad {\bf \Delta} = {\bf N}
+i\; {\bf M} \; ; \nonumber
\end{eqnarray}

\noindent it suffices  to  allow for the identities
\begin{eqnarray}
n_{0}  = \cos {\alpha \over 2} \; \mbox{ch}\;{\beta \over 2} \; ,
\qquad m_{0}  =  - \sin {\alpha \over 2} \; \mbox{sh}\;{\beta
\over 2}  \nonumber
\\
{\bf n} = \sin {\alpha \over 2}\; \mbox{ch}\; {\beta \over 2} \;
{\bf N} -
          \cos {\alpha \over 2}\; \mbox{sh}\; {\beta \over 2} \; {\bf M}
          \nonumber
          \\
\qquad {\bf m} = \cos {\alpha \over 2}\; \mbox{sh}\; {\beta \over
2} \; {\bf N} +
          \sin {\alpha \over 2}\; \mbox{ch}\; {\beta \over 2} \; {\bf M}   \; .
\nonumber
\\
\label{E}
\end{eqnarray}

The set of spinor matrices  (\ref{13a})  at any  fixed vector
${\bf \Delta}, \; {\bf \Delta }^{2}=1$ consists of a 2-parametric
subgroup with Abelian  group multiplication law:
\begin{eqnarray}
{\gamma '' \over 2} =  {\gamma  ' \over 2} + { \gamma \over 2}  \;
. \label{13b}
\end{eqnarray}

Complex rotation matrices $O$ in $(\gamma,
\Delta)$-parametrization look
\begin{eqnarray}
O =\left | \begin{array}{lll}
 1 - F (\Delta_{2}^{2} + \Delta_{3}^{2})   &
  -2 \sin \gamma \Delta_{3} + F \Delta_{1}\Delta_{2}    &
 +2 \sin \gamma \Delta_{2} + F \Delta_{1}\Delta_{3}
  \\
  +2 \sin \gamma \Delta_{3} + F \Delta_{1}\Delta_{2}
     &  1 - F  (\Delta_{3}^{2} + \Delta_{1}^{2})   &
       -2 \sin \gamma \Delta_{1} + F \Delta_{2}\Delta_{3}   \\
 -2 \sin \gamma \Delta_{2} + F \Delta_{1}\Delta_{3}
     &   +2 \sin \gamma \Delta_{1} +F \Delta_{2}\Delta_{3}   &  1 - F(\Delta_{1}^{2} + \Delta_{2}^{2})
 \end{array} \right |
 \nonumber
  \end{eqnarray}

\noindent where $F = 1 - \cos \gamma  $. We  need one simple
property of these 2-parametric subgroups (\ref{13b})
  an any fixed ${\bf \Delta}$  -- each of them leaves
invariant a definite  complex non-isotropic 3-vector, fixed up to
any non-zero  complex factor $\lambda$ (equation  (\ref{15a}) is verified by   direct
calculation)
\begin{eqnarray}
{\bf \Delta}^{2} = 1, \qquad O (\gamma, {\bf \Delta})\; \lambda \;
{\bf \Delta }  = \lambda \;  {\bf \Delta}
 \; ;
\label{15a}
\end{eqnarray}
\begin{eqnarray}
O_{1j} \;  \Delta_{j}=
 [\;
\Delta_{1} -  F (\Delta_{2}^{2} + \Delta_{3}^{2})\Delta_{1}
-\sin \gamma \;\Delta_{3}\Delta_{2}
+
\nonumber
\\
 +F  \Delta_{1}\Delta^{2}_{2}    +
     \sin \gamma \; \Delta_{2}
\Delta_{3}+
 F  \Delta_{1}\Delta_{3}^{2} \; ]=
 \; \Delta_{1} ,
 \nonumber
 \\[3mm]
O_{2j}  \;  \Delta_{j}= [\;
 \sin \gamma \; \Delta_{3}\Delta_{1}  + F  \Delta_{1}^{2} \Delta_{2}     +
 \Delta_{2}
-
\nonumber
\\
 -  F (\Delta_{3}^{2} + \Delta_{1}^{2})\Delta_{2} - \sin \gamma \;
\Delta_{1}\Delta_{3}  + F \Delta_{2}\Delta_{3}^{2} \; ] =
\Delta_{2}  ,
\nonumber
\\[3mm]
O_{3j}   \Delta_{j}=  [\;
 -\sin \gamma  \Delta_{2}\Delta_{1} + F \Delta_{1}^{2}\Delta_{3}     +   \sin \gamma
  \; \Delta_{1}\Delta_{2}
 +
 \nonumber
\\
 + (1-\cos \gamma)  \Delta_{2}^{2}\Delta_{3}
+  \Delta_{3}  - (1- \cos \gamma) (\Delta_{1}^{2} + \Delta_{2}^{2}) \Delta_{3} =  \Delta_{3}  .
\nonumber
\end{eqnarray}

Given arbitrary non-isotropic complex  vector  ${\bf K}$, to
construct a corresponding small subgroup in SO(3.C), it suffices
to have found a corresponding vector ${\bf \Delta}$  normalized on
+1. Let us detail  this task:
\begin{eqnarray}
{\bf K} =   {\bf n} + i \; {\bf m} = K \; {\bf  \Delta} \; ,
\qquad {\bf \Delta} = {\bf N} + i {\bf M}  \; , \nonumber
\\
 {\bf \Delta}^{2} =
  1\; , \qquad
{\bf N}^{2} - {\bf M}^{2} =1\;, \qquad 2i \; {\bf N} \; {\bf M}  =
0  \; ,\nonumber
\\
K^{2} =  ( {\bf n}^{2} - {\bf m}^{2}) +2i\; {\bf n}\; {\bf m}  =
I_{1} + i I_{2}\; ,\qquad I_{1},I_{2}= \mbox{inv} \;  , \nonumber
\end{eqnarray}

\noindent that is
\begin{eqnarray}
{\bf K} =  {\bf n} + i  \; {\bf m} = K \; {\bf  \Delta} =
 \pm \; \sqrt{({\bf n}^{2} - {\bf m}^{2}) +2i\; {\bf n}\; {\bf m}  }  \; \; ( {\bf N} + i {\bf M} )
\; . \label{17}
\end{eqnarray}

\noindent Complex invariant $K^{2}$  may be presented differently
\begin{eqnarray}
K^{2} =  ({\bf n} ^{2} - {\bf m}^{2}) +2i\; {\bf n}\; {\bf m}   =
  I  \; ( \cos 2 \mu  +  i\; \sin 2 \mu ) \; ,
  \nonumber
  \\
  I = + \sqrt{ ({\bf n}^{2} - {\bf m}^{2})^{2} + 4 ({\bf n}\; {\bf
m})^{2} }\; ,
\nonumber
\\
\cos 2\mu =  { I_{1 } \over \sqrt{I^{2}_{1} +I^{2}_{2}}} = { {\bf
n}^{2} - {\bf m}^{2} \over \sqrt{ ({\bf n}^{2} - {\bf m}^{2})^{2}
+ 4 ({\bf n}\; {\bf m})^{2} } } \; ,
\nonumber
\\
\sin 2\mu = { I_{2 } \over \sqrt{I^{2}_{1} +I^{2}_{2}}}  = { 2
{\bf n}\; {\bf m} \over \sqrt{ ({\bf n}^{2} - {\bf m}^{2})^{2} + 4
({\bf n}\; {\bf m})^{2} }} \; . \label{19}
\end{eqnarray}

\noindent Therefore,  the complex ${\bf K}$  may be written in the
form
\begin{eqnarray}
{\bf K} = {\bf n} + i  \; {\bf m}= \sqrt{I \; e^{2i\mu}} \;  ( {\bf N} + i
{\bf M} )
\nonumber
\\
=  [\;  ({\bf n}^{2} - {\bf m}^{2})^{2} + 4 ({\bf n}\; {\bf
m})^{2}\;  ]^{1/4}
  \;\;  e^{i\mu} \;\;   ( {\bf N} + i {\bf M} )\; ;
\nonumber
\end{eqnarray}

\noindent from whence one obtains an expression for   $ {\bf N} +
i {\bf M} $:
\begin{eqnarray}
 {\bf N} + i {\bf M} =
 {  e^{-i \mu}\; (  {\bf n} + i\; {\bf m}) \over
 [ ({\bf n}^{2} -{\bf m}^{2})^{2} + 4 ({\bf n}\; {\bf m})^{2} \; ]^{1/4}  } \; \; .
\label{20}
\end{eqnarray}

In two particular cases, these formulas are much simplified:

\vspace{3mm} $ \underline{ ( I_{1} \neq 0\; , \;\; I_{2}=0 ) } $
\begin{eqnarray}
I(a) \qquad {\bf n}^{2} > {\bf m }^{2}\; , \qquad \mu = 0 \; ,
 \nonumber
\\
{\bf n} + i \; {\bf m} =  \sqrt{ {\bf n}^{2} - {\bf m}^{2} } \; {
{\bf n} + i {\bf m} \over \sqrt{ {\bf n}^{2} - {\bf m}^{2}} }
\equiv \sqrt{ {\bf n}^{2} - {\bf m}^{2}}\; ( {\bf N} + i {\bf M} ) \; ,
 \nonumber
\\
\label{21a}
\\
 I(b) \qquad  {\bf n}^{2} < {\bf m}^{2} \; , \qquad \mu = {\pi \over 2} \; ,
  \nonumber
 \\
{\bf n} + i \; {\bf m} =   \sqrt{  {\bf n}^{2}  - {\bf m}^{2}  }
\; \;  {  (-i)({\bf n} + i {\bf m}) \over \sqrt{ -({\bf n}^{2} -
{\bf m}^{2}) } }  \equiv \sqrt{ {\bf n}^{2} - {\bf m}^{2} }\; (
{\bf N} + i {\bf M} ) \; ,
\nonumber
\\
\label{21b}
\end{eqnarray}

\begin{eqnarray}
\underline{ (I_{1} = 0\; , \qquad I_{2} \neq 0 ) } \; , \qquad
\qquad \cos 2\mu = 0 \; , \qquad \sin 2\mu = {  {\bf n}\; {\bf m}
\over \sqrt{   ({\bf n}\; {\bf m})^{2} }}  \; , \nonumber
\\
II(a) \qquad {\bf n}\; {\bf m} > 0 \; , \qquad \mu = { \pi \over
4}  \; ,\nonumber
\\
{\bf n} + i \; {\bf m} =  \sqrt{+ 2({\bf n} {\bf m})} \; {
e^{-i\pi/4} \; \; ({\bf n} + i {\bf m}) \over \sqrt{+ 2({\bf n}
{\bf m})} } \equiv \sqrt{+ 2({\bf n} {\bf m})} \; ( {\bf N} + i
{\bf M} ) \; ,
\nonumber
\\
\label{22a}
\\
II(b) \qquad {\bf n}\; {\bf m} > 0 \; , \qquad \mu = { \pi \over
4}  \; , \nonumber
\\
{\bf n} + i \; {\bf m} =  \sqrt{ 2({\bf n} {\bf m})} \; \; {
e^{-i3\pi/4} \; \; ({\bf n} + i {\bf m}) \over \sqrt{- 2({\bf n}
{\bf m})} } \equiv \sqrt{ 2({\bf n} {\bf m})} \; ( {\bf N} + i
{\bf M} )\; .
\nonumber
\\
\label{22b}
\end{eqnarray}

Turning back to the main relationship
\begin{eqnarray}
{\bf K}^{2} \neq 0\; , \; {\bf K} = K \; {\bf \Delta}\; ,
\qquad {\bf \Delta}^{2} = 1 \; , \nonumber
\\
O (\gamma, {\bf \Delta})\; \; K  {\bf \Delta}  = K  {\bf \Delta}
 \; , \;  {\bf \Delta}^{2}=1 \; ,
\label{23}
\end{eqnarray}

\noindent we note two special cases when the sense of the
parameter  $\gamma = \alpha + i \beta$ is evident in physical
terms:

\vspace{3mm} \underline{the first}
\begin{eqnarray}
{\bf \Delta}= (N + i M) \; {\bf e }\; , \qquad  \gamma = \alpha \;
,
\nonumber
\\
B =  \cos \alpha - i  \; \sin \alpha  \; {\bf e} \;
\vec{\sigma}  \; ,\nonumber
\\
O (\alpha, {\bf e})\;  (N + i M) {\bf e }   = (N + i M) \; {\bf e
} \; ,
\nonumber
\\
 O (\alpha, {\bf e}) \in SO(2)\; ; \label{24'}
\end{eqnarray}

\vspace{3mm} \underline{the second }
\begin{eqnarray}
{\bf \Delta}= (N + i M) \; {\bf e } \; , \qquad \gamma = i\;
\beta\; ,
\nonumber
\\
B =  \mbox{ch}\; \beta + \mbox{sh}\; \beta \;
{\bf e} \; \vec{\sigma}   \; , \nonumber
\\
O (i \beta, {\bf e})\;  (N + i M) \; {\bf e }  = (N + i M) {\bf e
} \; ,
\nonumber
\\
 O (\alpha, {\bf e}) \in SO(1,1)\; . \label{25}
\end{eqnarray}

\noindent In particular, the above variants  $I(a), I(b)$ are of
that type:
\begin{eqnarray}
I(a) \qquad {\bf m} = 0\;  , \; {\bf K} = {\bf n}
=  \sqrt{+{\bf n}^{2}} \;\; { {\bf n} \over  \sqrt{+{\bf n}^{2}} }\; ,
 \nonumber
\\
I(b) \qquad {\bf n} = 0\;  ,   \; {\bf K} = i\;
{\bf m}= \sqrt{-{\bf m}^{2}} \;\;  { {\bf m} \over  \sqrt{{\bf
m}^{2}} } \; .
\end{eqnarray}

\section{ On reduction of a complex non-isotropic vector to a
real form}

Let us demonstrate that the case of an arbitrary complex vector
${\bf \Delta}, \; {\bf \Delta}^{2}=1$ always can be reduced to a
real form by means of an appropriate Lorentz transformation. To
this end, let start with any complex vector of unit length:
\begin{eqnarray}
{\bf \Delta} = {\bf N} + i {\bf M} \; , \;   {\bf N}^{2} -
{\bf M}^{2} = 1 \; , \;  {\bf N} \; {\bf M}=0 \; , \nonumber
\\
{\bf N} =  \mbox{ch}\; \rho \; {\bf N}_{0}\; , \qquad  {\bf N}_{0}^{2}
= 1 \; ,
\nonumber
\\
{\bf M} =  \mbox{sh}\; \rho \; {\bf M}_{0}\; , \;
{\bf M}_{0}^{2} = 1 \; , \; {\bf N}_{0}\; {\bf M}_{0}  = 0 \; ;
\label{B.1}
\end{eqnarray}

\noindent and find  a matrix $S \in SO(3.C)$ satisfying equation
\begin{eqnarray}
 S ({\bf N} + i {\bf M}) = {\bf e}  + i  0 \;  , \;   {\bf e}^{2} = +1 \;  .
\label{B.2}
\end{eqnarray}

\noindent Eq. (\ref{B.2}) can be written differently
\begin{eqnarray}
{S + S^{*} \over 2} \; ({\bf N} + i {\bf M}) +  {S - S^{*} \over
2}  ({\bf N} + i {\bf M}) = {\bf e} + i 0  . \nonumber
\end{eqnarray}

\noindent With the notation
\begin{eqnarray}
{S + S^{*} \over 2} = R \; , \; {S - S^{*} \over 2} = -i \;
J\; , \; S = R -i J  \nonumber
\end{eqnarray}

\noindent we get two equations
\begin{eqnarray}
R \; {\bf N} + J\; {\bf M} = {\bf K}_{0} \; , \qquad R \; {\bf M}
- J\; {\bf N} = 0  \; ; \nonumber
\end{eqnarray}

\noindent
 they can be written as
 \begin{eqnarray}
R \;  \mbox{ch}\; \rho \; {\bf N}_{0}  + J\; \mbox{sh}\; \rho \;
{\bf M}_{0}  = {\bf e} \; ,
\nonumber
\\
 R \; \mbox{sh}\; \rho \; {\bf
M}_{0} = + J\; \mbox{ch}\; \rho \; {\bf N}_{0}  \; . \label{B.4}
\end{eqnarray}

\noindent Second relation in  (\ref{B.4})  is equivalent to
\begin{eqnarray}
J^{-1} R \; \mbox{th}\; \rho \; {\bf M}_{0} = + \;  {\bf N}_{0} \;
. \label{B.5}
\end{eqnarray}

\noindent  However, an orthogonal  rotation $O_{1}=O({\bf
c}_{1})$, changing a unite length  vector
 ${\bf M}_{0}$  into another vector unit length vector
${\bf N}_{0} $  is well known  \cite{Fedorov-1980}
\begin{eqnarray}
J^{-1} R \; \mbox{th}\; \rho = O_{1} \; ,
\qquad
  O _{1}\; {\bf
M}_{0} = {\bf N}_{0} \;, \qquad  {\bf c}_{1}   = { {\bf M}_{0}
\times {\bf N}_{0} \over 1 + {\bf M}_{0} \; {\bf N}_{0}} \; .
\label{B.6}
\end{eqnarray}

\noindent Substituting  this  into the first equation in
(\ref{B.4}) we get
\begin{eqnarray}
\mbox{sh}\; \rho \; ( R \;   J^{-1} R    + J   )\; {\bf M}_{0}  =
{\bf e} \; . \label{B.7}
\end{eqnarray}

\noindent Rotation transforming the vector  ${\bf M}_{0}$  into
${\bf e}$  (note it as  $O_{2}$ ) is
\begin{eqnarray}
\mbox{sh}\; \rho \; ( R \;   J^{-1} R    + J   ) = O_{2} \; ,
\qquad
O _{2}\; {\bf M}_{0}  = {\bf e} , \;\;
 {\bf c}_{2} = {
{\bf M}_{0} \times {\bf e} \over 1 + {\bf M}_{0} \; {\bf e}} \; .
\label{B.8}
\end{eqnarray}

\noindent Therefore, we know expressions for two matrices  $O_{1}$
and $O_{2}$ --  see (\ref{B.6}) and   (\ref{B.8}),   in terms of
which two other
 $R$ and $J$ are given:
\begin{eqnarray}
J^{-1} R  = {O_{1} \over  \mbox{th}\; \rho} \; , \qquad R \;
J^{-1} R    + J    = {O_{2}  \over \mbox{sh}\; \rho} \; .
\label{B.9}
\end{eqnarray}

\noindent Solving eqs.  (\ref{B.9}) is quite elementary:
\begin{eqnarray}
 R = {J \; O_{1}  \over  \mbox{th}\; \rho} \; ;
\nonumber
\end{eqnarray}

\noindent and substitution this $R$  into second equation in
(\ref{B.9}) we get
\begin{eqnarray}
  {J \;O_{1}  \over  \mbox{th}\; \rho}  \;   J^{-1} {J \; O_{1}  \over  \mbox{th}\; \rho}
     + J    = {O_{2}  \over  \mbox{sh}\; \rho } \; , \qquad
J  \; (\mbox{ch}^{2}\rho \; O_{1}^{2} + \mbox{sh}^{2}\rho)      =
\mbox{sh}\; \rho   \; O_{2} \; . \label{B.10}
\end{eqnarray}

\noindent
 Thus $J$ and  $R$ have been found:
 \begin{eqnarray}
J = \mbox{sh}\; \rho \; O_{2} \; (\mbox{ch}^{2}\rho \; O_{1}^{2} +
\mbox{sh}^{2}\rho) ^{-1}  \; , \nonumber
\\
 R = \mbox{ch}\; \rho \; O_{2}
\; (\mbox{ch}^{2}\rho \; O_{1}^{2} + \mbox{sh}^{2}\rho) ^{-1} \;
O_{1}  \; ; \label{B.11}
\end{eqnarray}

\noindent correspondingly, the the $S$ transformation we need is
\begin{eqnarray}
S = R - i J\; \in SO(3.C)  \; , \nonumber
\\
S =  O_{2} \; (\mbox{ch}^{2}\rho \; O_{1}^{2} + \mbox{sh}^{2}\rho)
^{-1} \; [ \mbox{ch}\; \rho \;  \; O_{1} -i \; \mbox{sh}\; \rho \;
]\;.
\label{B.10'}
\end{eqnarray}

One may note one special case to choose the vector ${\bf e}$.
Indeed, let it be
 ${\bf e} = {\bf
N}_{0}, \; O_{2} = O_{1} $  which leads to
\begin{eqnarray}
J = \mbox{sh}\; \rho \; O_{1} \; (\mbox{ch}^{2}\rho \; O_{1}^{2} +
\mbox{sh}^{2}\rho) ^{-1}  \; , \nonumber
\\[3mm]
R = \mbox{ch}\; \rho \; O_{1} \; (\mbox{ch}^{2}\rho \; \Pi^{2} +
\mbox{sh}^{2}\rho) ^{-1} \; O_{1}  \; . \label{B.12}
\end{eqnarray}

Besides, one may choose the variant   ${\bf e} = {\bf M}_{0},
O_{2} = I$, then we arrive at
\begin{eqnarray}
J = \mbox{sh}\; \rho \;  (\mbox{ch}^{2}\rho \; O_{1}^{2} +
\mbox{sh}^{2}\rho) ^{-1}  \; , \nonumber
\\[3mm]
R = \mbox{ch}\; \rho \; \; (\mbox{ch}^{2}\rho \; O_{1}^{2} +
\mbox{sh}^{2}\rho) ^{-1} \; O_{1} \; . \label{B.13}
\end{eqnarray}

Let us turn again to  the stationary subgroup problem:
\begin{eqnarray}
{\bf K} = {\bf n} + i \; {\bf m} = K\; {\bf \Delta} \; , \;
{\bf \Delta}^{2} = 1  \; , \nonumber
\\
K = \sqrt{I_{1} + iI_{2}} \; , \; I_{1} = {\bf
n}^{2} - {\bf m}^{2} \; ,\; I_{2} = 2i\; {\bf n} {\bf m}
 \nonumber
\\[5mm]
O( \gamma, {\bf \Delta}) \; \sqrt{I_{1} + iI_{2}}\; {\bf \Delta} =
\sqrt{I_{1} + iI_{2}} \; {\bf \Delta} \; , \;\;  \Longrightarrow
\nonumber
\\
O( \gamma, {\bf N} +  i {\bf N}) \; ( {\bf n} + i {\bf m})  = (
{\bf n} + i {\bf m})\; ,
\label{B14b}
\end{eqnarray}

\noindent where
\begin{eqnarray}
{\bf n} + i{\bf m} =  \sqrt{I_{1} + iI_{2}} \; ( {\bf N} + i \;
{\bf M} )\; . \label{B.14c}
\end{eqnarray}

\noindent Therefore,  the main stationary  equation in an
arbitrary non-isotropic case
 may be written as
\begin{eqnarray}
O  ( \gamma,  {\bf \Delta} = { {\bf n} + i{\bf m} \over
\sqrt{I_{1} + iI_{2}} }  ( {\bf n} + i {\bf m}) = ( {\bf n} + i
{\bf m})  . \label{B.14d}
\end{eqnarray}

In turn, with the help of additional Lorentz transformation $S$
according to (\ref{B.2}), one may reduce equation (\ref{B.14d}) to
the form
\begin{eqnarray}
SO( \gamma, {\bf \Delta})S^{-1} K\; S \;{\bf \Delta} = K\; S{\bf
\Delta} \; =  K\; ({\bf e} + i \; 0 )  \nonumber
\end{eqnarray}

\noindent and further, with the use of the known identity in the
theory of the rotation group \cite{Fedorov-1980}, we arrive at the
basic relationship with clear  interpretation for $\gamma$ -- see
(\ref{24'})-(\ref{25}):
\begin{eqnarray}
O( \gamma, {\bf e} ) \sqrt{I_{1} + i I_{2}}  \; {\bf e}  =
\sqrt{I_{1} + i I_{2}}   \; {\bf e}  \;  , \;
 {\bf e}^{2} = 1\; .
\label{B.15}
\end{eqnarray}

In particular, the vector ${\bf e}$ may be taken as ${\bf e} =
{\bf N}_{0}$ or
 ${\bf e} = {\bf M}_{0}$.
 So, the values of invariants, $I_{1}$ and $I_{2}$, govern
 the possible most simple form for commutative parameters ${\bf n}'$ and ${\bf m}'$
 in different reference frames.
\begin{eqnarray}
[x_{a}, x_{b} ]_{-} = i\; \theta_{ab}\; , \qquad  \theta_{ab} \sim
({\bf n} + i {\bf  m} )\; ,
 \nonumber
\\[3mm]
\; [x'_{a}, x'_{b} ]_{-} = i\; \theta'_{ab}\; , \qquad
\theta'_{ab} \sim ({\bf n}' + i {\bf  m}' )  = \sqrt{I_{1} + i
I_{2}}\;  {\bf e} \; .
 \label{B.16}
\end{eqnarray}

\section{ On physical meaning of  2-parametric subgroup  \\  $O(\gamma = \alpha + i \beta,
{\bf \Delta})$ at arbitrary reference frame}

To have interpreted the complex parameter  $\gamma = \alpha + i
\beta$  of the subgroup  $O(\gamma = \alpha + i \beta, {\bf
\Delta})$  at arbitrary reference frame, let us decompose the
corresponding spinor elements into product of Euclidean rotation
and Lorentz boost:
\begin{eqnarray}
\cos {\alpha + i \beta \over 2} -i \sin {\alpha + i \beta \over 2}
\; ({\bf N} + i {\bf M}) \;\vec{\sigma} \nonumber
\\
=
 (\cos {a \over 2}  - i \sin {a \over 2}  {\bf a} \; \vec{\sigma})\;
(\mbox{ch}\; {b \over 2} + \mbox{sh}  {b \over 2}\; {\bf b}\;
\vec{\sigma} )\; ; \label{A.3}
\end{eqnarray}

\noindent it suffices to solve the problem (it is a
spinor variant of the well-known problem of factorization of any
Lorentz matrix into rotation and boost -- see, for instance, in
\cite{Fedorov-1980}):
\begin{eqnarray}
k_{0} + {\bf k} \; \vec{\sigma} = ( a _{0} -i {\bf a} \;
\vec{\sigma}  ) (b_{0} + {\bf b} \; \vec{\sigma}) \nonumber
\\
 =(a_{0}b_{0} - i\; {\bf  a} \; {\bf b} ) +
 (\; a_{0} \; {\bf b}  + {\bf a} \times {\bf b}  - i \; b_{0} \; {\bf a}   \;) \; \vec{\sigma}\; ,
\nonumber
\\
k_{0}^{*} +  {\bf k}^{*} \vec{\sigma} = ( a_{0} +i {\bf a}
\vec{\sigma}  ) (b_{0} + {\bf b} \vec{\sigma})
 \nonumber
 \\
 = (a_{0}b_{0} + i\; {\bf  a} \; {\bf b} ) +
 (\; a _{0}\; {\bf b}  + {\bf a} \times {\bf b}  + i \; b_{0} \; {\bf a}   \;) \; \vec{\sigma} \; ;
\nonumber
\label{A.4}
\end{eqnarray}

\noindent which is equivalent to the system
\begin{eqnarray}
k_{0} = (a_{0}b_{0} - i\; {\bf  a} \; {\bf b} )  \; , \qquad
k_{0}^{*}  = (a_{0}b_{0} + i\; {\bf  a} \; {\bf b} ) \; ,\nonumber
\\
 {\bf k} = (\; a_{0} \; {\bf b}  + {\bf a} \times {\bf b}  - i
\; b_{0} \; {\bf a}   \;) \; ,
\nonumber
\\  {\bf k}^{*}  = (\; a_{0} \;
{\bf b}  + {\bf a} \times {\bf b}  + i \; b_{0} \; {\bf a} \;) \; ,
 \nonumber
\end{eqnarray}

\noindent or
\begin{eqnarray}
{k_{0} + k_{0}^{*}  \over 2 } = a_{0}b _{0}\; , \qquad {k_{0} -
k_{0}^{*} \over 2i}  = - \; {\bf  a} \; {\bf b} \; , \nonumber
\\
{ {\bf k}  + {\bf k}^{*}  \over 2  }  =  a _{0}\; {\bf b}  +
{\bf a} \times {\bf b}  \; , \; { i {\bf k}  - i {\bf k}^{*}
\over 2}  =   b _{0} {\bf a}  \; . \label{A.5}
\end{eqnarray}

\noindent With additional restrictions:
\begin{eqnarray}
a^{2}_{0} + {\bf a}^{2} = +1 \; , \qquad a _{0}= \pm \; \sqrt{1 -
{\bf a}^{2}} \; , \nonumber
\\
b^{2}_{0} - {\bf b}^{2} = +1 \; , \qquad  b_{0}  = + \sqrt{1 +{\bf
b}^{2}}  \geq +1  \label{A.6}
\end{eqnarray}

\noindent eqs.   (\ref{A.5}) take the form
\begin{eqnarray}
n_{0}  = \pm \; \sqrt{1 - {\bf a}^{2}} \; \; \sqrt{1 +{\bf b}^{2}}
\; , \qquad m_{0}  = - \; {\bf  a} \; {\bf b} \; , \nonumber
\\
 {\bf m}  =  \pm \; \sqrt{1 - {\bf a}^{2}}  \; {\bf b}  + {\bf a} \times {\bf b}  \; , \;\;\;
{\bf n}  =   \sqrt{1 +{\bf b}^{2}} \;  \; {\bf a}  \; .
\nonumber
\end{eqnarray}

\noindent From whence it follows
\begin{eqnarray}
n_{0}  = \pm \; \sqrt{1 - {\bf a}^{2}} \; \; \sqrt{1 +{\bf b}^{2}}
\; , \;  m_{0} = - \; {\bf  a} \; {\bf b} \; , \nonumber
\\
{{\bf m} \over n_{0}} =  {{\bf b} \over \sqrt{1 + {\bf b}^{2}} } +
{ {\bf a}  \over \pm \sqrt{1 -{\bf a}^{2}}} \times  {{\bf b} \over
\sqrt{1 + {\bf b}^{2}} }   \; ,
\nonumber
\\
 {{\bf n} \over n_{0}} =   {
{\bf a}   \over  \pm \sqrt{1 -{\bf a}^{2}}  }\;   .
\label{A.8}
\end{eqnarray}

\noindent With the help of variables ${\bf A},{\bf B}$:
\begin{eqnarray}
{{\bf b} \over \sqrt{1 + {\bf b}^{2}} } = {\bf B} \; ,
  b_{0}= \sqrt{1+{\bf b}^{2} }=
 {1 \over \sqrt{{1 - \bf B}^{2}}} \;,\;
{\bf b} = { {\bf B} \over \sqrt{ 1-{\bf B}^{2} } } \; ,
 \nonumber
\\
{{\bf a} \over \pm \sqrt{1 - {\bf a}^{2}} } =  \pm  {\bf A}\; ,
 \;\;
a_{0} =  \pm \sqrt{1 -{\bf a}^{2}}  = {1 \over \pm \sqrt{1 + {\bf
A}^{2}} }\;, \;\;
 {\bf a} = { {\bf A} \over  \sqrt{ 1 +{\bf A}^{2}
} } \; , \nonumber
\end{eqnarray}

\noindent  we get
\begin{eqnarray}
n_{0}  = \pm \; {1 \over \sqrt{1 + {\bf A} ^{2}}} \; {1 \over
\sqrt{1-{\bf B}^{2} }} \; , \qquad m_{0}  = - \; {{\bf  A}  \over
\sqrt{1 + {\bf A}^{2}}}  \; {{\bf  B}  \over \sqrt{1-{\bf B}^{2}}}\; ,
 \nonumber
\\
{ {\bf m} \over    n_{0} }  =  {\bf B}   +
 {\bf A}   \times  {\bf B}  \; , \qquad \qquad
{ {\bf n}   \over n_{0}}  =    {\bf A}  \;  . \label{A.10}
\end{eqnarray}

\noindent The vector  ${\bf B}$ may be resolved into a linear
combination
$
{\bf B} = \nu \; {\bf n} + \mu   \;  {\bf m} +  \sigma  \; {\bf n}
\times  {\bf m}  $
 which must obey
\begin{eqnarray}
{ {\bf m} \over    n_{0} }  =  \nu \; {\bf n} + \mu   \;  {\bf m}
+  \sigma  \; {\bf n} \times  {\bf m}    + { {\bf n}   \over
n_{0}}   \times  ( \nu \; {\bf n} + \mu   \;  {\bf m} +  \sigma \;
{\bf n} \times  {\bf m})   \nonumber
\end{eqnarray}

\noindent or
\begin{eqnarray}
{ {\bf m} \over    n_{0} }  =  \nu \; {\bf n} + \mu   \;  {\bf m}
+  \sigma  \; {\bf n} \times  {\bf m}    + { \mu    \over n_{0}}
\;  {\bf n}  \times   {\bf m} + {  \sigma  \over n_{0}} ({\bf
n}{\bf m}) \; {\bf n}  - {  \sigma  \over n_{0}} \; ({\bf n}^{2} )
\;  {\bf m}\;   \; . \nonumber
\end{eqnarray}

\noindent Therefore, we have the system
\begin{eqnarray}
\nu + {  \sigma  \over n_{0}} ({\bf n}{\bf m})= 0 \; , \qquad {1
\over n_{0}} = \mu - {  \sigma \; ({\bf n}^{2} )   \over n_{0}} \;
, \qquad \sigma + { \mu    \over n_{0}} = 0 \nonumber
\end{eqnarray}

\noindent with evident solution
\begin{eqnarray}
\sigma = -{1 \over n_{0}^{2} + {\bf n}^{2}}\;, \;\;  \mu = {n_{0}
\over n_{0}^{2} + {\bf n}^{2}}\;, \;\;  \nu =  { ({\bf n} {\bf m})
\over n_{0}} \; {1 \over n_{0}^{2} + {\bf n}^{2}}= -m_{0} \; {1
\over n_{0}^{2} + {\bf n}^{2}}\; . \nonumber
\end{eqnarray}

\noindent Thus, the factorization we need is found:
\begin{eqnarray}
k_{0} + {\bf k} \; \vec{\sigma} = ({ n_{0} -i {\bf n} \;
\vec{\sigma} \over \sqrt{n_{0}^{2} + {\bf n}^{2}} })\; ( {1 \over
\sqrt{1-{\bf B}^{2}}} + { {\bf B} \; \vec{\sigma} \over  \sqrt{
1-{\bf B}^{2} } })  \; ,
\nonumber
\\
{\bf B}= {   n_{0}    \;  {\bf m} - m_{0}  \; {\bf n}  +  \; {\bf
m} \times  {\bf n} \over  n_{0}^{2} + {\bf n}^{2}}\;.
 \label{A.13}
\end{eqnarray}

The problem of factorization may be solved easily with  opposite
order:
\begin{eqnarray}
k_{0} + {\bf k} \; \vec{\sigma} =  (b_{0} + {\bf b} \;
\vec{\sigma}) ( a_{0} -i {\bf a} \; \vec{\sigma}  )
 =(a_{0}b_{0} - i\; {\bf  a} \; {\bf b} ) +
 (\; a_{0} \; {\bf b}  - {\bf a} \times {\bf b}  - i \; b_{0} \; {\bf a}   \;) \; \vec{\sigma}\; ,
\nonumber
\\[3mm]
k_{0}^{*} +  {\bf k}^{*} \vec{\sigma} =  (b_{0} + {\bf b}
\vec{\sigma}) ( a_{0} +i {\bf a} \vec{\sigma}  )
 = (a_{0}b_{0} + i\; {\bf  a} \; {\bf b} ) +
 (\; a_{0} \; {\bf b}  - {\bf a} \times {\bf b}  + i \; b _{0}\; {\bf a}   \;) \; \vec{\sigma} \; ;
\label{A.19}
\end{eqnarray}

\noindent it  reduces to the system (in comparison with
(\ref{A.4}) only the sign at the vector product has been changed
on opposite)
\begin{eqnarray}
k_{0} = (a_{0}b_{0} - i\; {\bf  a} \; {\bf b} )  \; , \qquad
k_{0}^{*}  = (a_{0}b_{0} + i\; {\bf  a} \; {\bf b} ) \; ,\nonumber
\\
 {\bf k} = (\; a _{0}\; {\bf b}  - {\bf a} \times {\bf b}  - i
\; b_{0} \; {\bf a}   \;) \; , \qquad   {\bf k}^{*}  = (\; a
_{0}\; {\bf b}  - {\bf a} \times {\bf b}  + i \; b_{0} \; {\bf a}
\;) \; , \nonumber
\end{eqnarray}

\noindent or
\begin{eqnarray}
n_{0} = a_{0}b_{0} \; , \qquad m_{0} = - \; {\bf  a} \; {\bf b} \; ,
\qquad
 {\bf m}   =  a_{0} \; {\bf b}  - {\bf a} \times {\bf b}  \; , \qquad
 {\bf n}  =   b _{0}\; {\bf a}  \; .
\label{A.20}
\end{eqnarray}

\noindent Further analysis is the same, the final result is
\begin{eqnarray}
k_{0} + {\bf k} \; \vec{\sigma} = ( {1 \over \sqrt{1-{\bf B}^{2}}}
+ { {\bf B} \; \vec{\sigma} \over  \sqrt{ 1-{\bf B}^{2} } }) \; ({
n_{0} -i {\bf n} \; \vec{\sigma} \over \sqrt{n_{0}^{2} + {\bf
n}^{2}} }) \; ,
\nonumber
\\
{\bf B}= {   n_{0}    \;  {\bf m} - m_{0}  \; {\bf n}  -  \; {\bf
m} \times  {\bf n} \over  n_{0}^{2} + {\bf n}^{2}}\; .
 \label{A.22}
\end{eqnarray}

The factorizations produced can be translated to parameters
$(\gamma/2 , {\bf \Delta})$:
\begin{eqnarray}
B = k_{0} + {\bf k} \; \vec{\sigma} =  \cos {\alpha + i \beta
\over 2} -i \sin  {\alpha + i \beta \over 2} \; {\bf \Delta} \; ,
\qquad {\bf \Delta} = {\bf N} + i {\bf M}
\nonumber
\end{eqnarray}

\noindent with the help of the formulas (\ref{E}).

\section{ The problem of a small group in  $SO(3,C)$,   isotropic case
} \vspace{5mm}

Let us consider transformations of the group $SL(2.C)$ with
isotropic vector
 ${\bf k}$:
\begin{eqnarray}
k_{0}= \pm 1\; , \qquad B =  \pm \;( I +\; {\bf k}  \;
\vec{\sigma} ) \; , \qquad
{\bf k}^{2} =  0\; ,
\nonumber
\\
 I_{1}=  {\bf n}^{2} - {\bf m}^{2} = 0 \;
, \qquad  I_{2}=
  +2 \;  {\bf n} {\bf m} = 0 \; .
 \label{22}
 \end{eqnarray}

Evidently, vectors ${\bf k}$ are fixed within arbitrary complex
numerical factor
 $ {\bf k} ' =  z\; {\bf \Delta}, \;{\bf \Delta}^{2} = 0 $, therefore one may
 construct the following 2-parametric subgroups in $SL(2.C)$;
bellow we are interested mainly in corresponding elements in the
 $SO(3.C)$ group when the factor $\delta = \pm 1$, has no effect:
\begin{eqnarray}
\delta ' (I +  z '\; {\bf k }  \; \vec{\sigma}  )\; \; \delta  (I
+  z \; {\bf k}  \; \vec{\sigma}  ) = \delta' \delta\; [\; I +  (z
' + z) \; {\bf k}  \; \vec{\sigma} \; ]  \label{25a}
\end{eqnarray}

\noindent and  correspondingly
$
O(z {\bf k}) \; {\bf k} = {\bf k} \; , \; {\bf k}^{2} = 0
$
  where
\begin{eqnarray}
O(z {\bf k}) = \left | \begin{array}{rrr}
 1 + 2 z^{2}(k_{2}^{2} + k_{3}^{2})   &   -2 zi k_{3} - 2z^{2}k_{1}\Delta_{2}    &
 +2 z i k_{2} - 2z^{2}k_{1}\Delta_{3}  \\
 +2zi k_{3} - 2 z^{2} k_{1}k_{2}     &  1 +2 z^{2}(k_{3}^{2} + k_{1}^{2})   &
  -2 zik_{1} -  2z^{2}k_{2}k_{3}   \\
 -2 zi k_{2} - 2z^{2}k_{1}k_{3}     &   +2zik_{1} -
  2z^{2}k_{2}k_{3}    &  1 +2z^{2} (k_{1}^{2} + k_{2}^{2})
 \end{array} \right | .
\nonumber
 \end{eqnarray}

\noindent Formulas  are much simplified in particular cases:
\begin{eqnarray}
{\bf k } =  (  k_{1},k_{2},0) \; , \qquad k_{1}^{2}  +  k_{2}^{2}
= 0   \; , \nonumber
\\
\left | \begin{array}{rrr}
 1 +2 z^{2}k_{2}^{2}    &   - 2z^{2}k_{1}k_{2}    &  +2 z i k_{2}   \\
 - 2z^{2}k_{1}k_{2}     &  1 +2 z^{2} k_{1}^{2}   &  -2i  zk_{1}    \\
 -2i zk_{2}     &   +2i zk_{1}  &  1 + 2z^{2} (k_{1}^{2} + k_{2}^{2})
 \end{array} \right |
 \left | \begin{array}{c}
 k_{1} \\ k_{2}\\ 0
 \end{array} \right | =
 \left | \begin{array}{c}
 k_{1} \\ k_{2}\\ 0
 \end{array} \right |\; ,
\nonumber
\end{eqnarray}
\begin{eqnarray}
{\bf k } =  ( 0,  k_{2},k_{3}) \; , \qquad k_{1}^{1}  +  k_{2}^{2}
= 0  \; , \nonumber
\\
\left | \begin{array}{rrr}
 1 +2 z^{2}(k_{2}^{2} + k_{3}^{2})   &   -2i z k_{3} & +2 iz k_{2}  \\
 +2iz k_{3}     &  1 +2 z^{2}k_{3}^{2}  &  - 2z^{2}k_{2}k_{3}   \\
 -2 izk_{2}     &  -  2z^{2}k_{2}k_{3}    &  1 +2z^{2} k_{2}^{2}
 \end{array} \right |
 \left | \begin{array}{c}
 0 \\ k_{2} \\ k_{3}
 \end{array} \right | =
 \left | \begin{array}{c}
 0 \\ k_{2} \\ k_{3}
 \end{array} \right |\; ,
\nonumber
\end{eqnarray}
\begin{eqnarray}
{\bf k } =  (  k_{1},0, k_{3}) \; , \qquad k_{1}^{2}  +  k_{3}^{2}
= 0  \; , \nonumber
\\
  \left | \begin{array}{rrr}
 1 +2 z^{2} k_{3}^{2}   &   -2i z k_{3}    &  -2z^{2}k_{1}k_{3}  \\
 +2iz k_{3}      &  1 +2 z^{2}(k_{3}^{2} + k_{1}^{2})  &  -2i zk_{1}    \\
  - 2z^{2}k_{1}k_{3}     &   +2izk_{1}    &  1 +2z^{2} k_{1}^{2}
 \end{array} \right |
 \left | \begin{array}{c}
  k_{1} \\ 0 \\ k_{3}
 \end{array} \right |=
 \left | \begin{array}{c}
 k_{1} \\ 0 \\ k_{3}
 \end{array} \right | \; .
\label{26}
\end{eqnarray}

To reach some base to interpret the   complex parameter $z$ in
physical terms, we should use the corresponding $4\times4$ Lorentz
matrices $L(\pm(1, -i{\bf n} + {\bf m}))$. The $z = \lambda
e^{i\sigma}$ - freedom in vector ${\bf k}$ is described by
relation ${\bf k}' = \lambda  \; e^{i\sigma} \; {\bf k}\; $:
\begin{eqnarray}
(-i {\bf n}' +  \; {\bf m}' ) = \lambda \; ( \cos \sigma + i\;
\sin \sigma ) \; (-i {\bf n} +  {\bf m}) \; , \nonumber
\\
{\bf n}' = \lambda \; ( \cos \sigma \; {\bf n} - \sin \sigma \;
{\bf m} )\; , \qquad {\bf m}' = \lambda (\sin \sigma \; {\bf n} +
\cos \sigma \; {\bf m} )  \; , \nonumber
\\
{\bf n}^{'2}= \lambda ^{2} \; {\bf n}^{2} \; , \qquad {\bf
m}^{'2}=   \lambda ^{2} \; {\bf m}^{2} \; ,\nonumber
\\
 {\bf n}'\; {\bf
m}' = 0 \;, \qquad {\bf n}' \times {\bf m}' = \lambda ^{2} \; {\bf
n} \times {\bf m} \;    . \label{30}
\end{eqnarray}

To have additional ground to interpret physically the parameter
$z$, let us factorized spinor matrix  $ B(\pm (1, -i {\bf n} +
{\bf m}))$ into the product of rotation and boost
\begin{eqnarray}
a_{0}^{2} + {\bf a}^{2} = 1 \; , \qquad  b_{0}^{2} - {\bf b}^{2} =
1  \; , \nonumber
\\
\pm \; [\; I  + (- i  {\bf n} +  {\bf m}) \; \vec{\sigma} \; ] =
(a_{0} -i\;  {\bf a}\;  \vec{\sigma})\; (b_{0} + {\bf b} \;
\vec{\sigma}) \; ,\nonumber
\\
= a_{0}b_{0} + a_{0}\;  {\bf b} \; \vec{\sigma}  -i b_{0}\;  {\bf
a} \; \vec{\sigma} -i \; [\;   {\bf a} {\bf  b}
 + i\;  ({\bf a}\times {\bf b}) \; \vec{\sigma}\;  ] \; .
\label{37}
\end{eqnarray}

\noindent The problem is reduced to the system
\begin{eqnarray}
 {\bf a} \; {\bf  b}=0 \; , \qquad   a_{0}\; b_{0}  = \pm \; 1 \; ,
\nonumber
\\
\pm \; {\bf n}=   b_{0}\;  {\bf a}  \; , \qquad \Longrightarrow
\qquad     {\bf a}  = a_{0} \; {\bf n}  \; ,\nonumber
\\
\pm \; {\bf m} = a_{0}\;  {\bf b} + ({\bf a}\times {\bf b})\; ,
\qquad \Longrightarrow  \qquad b_{0}\; {\bf m} =  {\bf b}  + {\bf
n} \times  {\bf b}\; . \label{38}
\end{eqnarray}

\noindent One can resolve the  vector  ${\bf b}$  into the linear
combination
$
{\bf b} =  b_{0} \; [\;  \alpha \; {\bf n} + \beta \; {\bf m} +
\gamma \;( {\bf n} \times {\bf m} ) \; ]
$,
from whence it follows
\begin{eqnarray}
{\bf m} = \alpha \; {\bf n} + \beta \; {\bf m} + \gamma \;( {\bf
n} \times {\bf m} )  +
  \beta \; {\bf n} \times  {\bf m} - \gamma \; n^{2} \; {\bf m} \; ,
\nonumber
\end{eqnarray}

\noindent that is
\begin{eqnarray}
{\bf b} =  b_{0}  \; { ({\bf m} -  {\bf n} \times {\bf m} ) \over
1 + n^{2}} \; . \label{39}
\end{eqnarray}

\noindent Thus, the  factorization has been found:
\begin{eqnarray}
\pm \; [\; I - i ( {\bf n} + i \; {\bf m}) \; \vec{\sigma} \; ] =
(a_{0} -i\; a_{0}\; {\bf n}\;  \vec{\sigma})\;\;  ( \; b_{0} +
b_{0}  \; { ({\bf m} -  {\bf n} \times {\bf m} )   \over  1 +
n^{2}} \; \vec{\sigma} \;   ) \; ,  \nonumber
\\
a_{0}^{2} + a_{0}^{2} \; n^{2} = 1 \; , \;\;  \Longrightarrow \;\;
a_{0} =  \pm\; {1 \over  \sqrt{1 + n^{2}}  }  \; ,\nonumber
\\
b_{0}^{2} - b_{0}^{2} { ({\bf m} -  {\bf n} \times {\bf m} )^{2}
\over  (1 + n^{2})^{2} }= 1 \; , \;\;  \Longrightarrow  \;\; b_{0}
=  \; \sqrt{1 +n^{2} } \; . \label{40}
\end{eqnarray}

In the same manner, one solves  the problem with opposite order:
\begin{eqnarray}
\pm \; [\; I - i ( {\bf n} + i \; {\bf m}) \; \vec{\sigma} \; ] =
(b_{0} + {\bf b} \; \vec{\sigma}) \;\; (a_{0} -i\;  {\bf a}\;
\vec{\sigma})  \; , \nonumber
\\
= a_{0}b_{0} + a_{0}\;  {\bf b} \; \vec{\sigma}  -i b_{0}\;  {\bf
a} \; \vec{\sigma} -i \; [\;   {\bf a} {\bf  b}
 - i\;  ({\bf a}\times {\bf b}) \; \vec{\sigma}\;  ] \; ,
\label{41}
\end{eqnarray}

\noindent which results in
\begin{eqnarray}
\pm \; [\; I - i ( {\bf n} + i \; {\bf m}) \; \vec{\sigma} \; ]
= (a_{0} -i\; a_{0}\; {\bf n}\;  \vec{\sigma})\; ( \; b_{0} +
b_{0}  \; { ({\bf m} +  {\bf n} \times {\bf m} )   \over  1 +
n^{2}} \; \vec{\sigma} \;   ) \; ,  \nonumber
\\
a_{0}^{2} + a_{0}^{2} \; n^{2} = 1 \; , \;\;  \Longrightarrow \;\;
a_{0} =  \pm\; {1 \over  \sqrt{1 + n^{2}}  }  \; ,  \nonumber
\\
b_{0}^{2} - b_{0}^{2} { ({\bf m} +  {\bf n} \times {\bf m} )^{2}
\over  (1 + n^{2})^{2} }= 1 \; , \;\;  \Longrightarrow  \;\; b_{0}
=  \; \sqrt{1 +n^{2} } \; . \label{42}
\end{eqnarray}

The $z = \lambda e^{i\sigma}$ -- freedom in  ${\bf k}$ plays
essential role in the factorizations:
\begin{eqnarray}
\pm  [ I  + (- i  {\bf n}' + {\bf m}') \; \vec{\sigma} \; ] =
(a_{0}' -i  a_{0}' {\bf n}'\;  \vec{\sigma})  \;  ( b_{0}' +
b_{0}'   { ({\bf m}' -  {\bf n}' \times {\bf m} ')
 \over  1 + n^{'2}}  \vec{\sigma}    ) \; ,
\nonumber
\\
 a_{0}' =   \pm\; {1 \over  \sqrt{1 +  \lambda^{2} \; n^{2}}  } \; ,
\;\;
 b_{0}' =    {1 \over  \sqrt{1 +  \lambda^{2} \; n^{2}}  } \;
 \; , \;\;
  {\bf n}' \times {\bf m}' =
\lambda^{2}  \; {\bf n} \times {\bf m} \; ; \nonumber
\\
\label{43}
\end{eqnarray}

\noindent and
\begin{eqnarray}
\pm  [\; I  + (- i   {\bf n}' +  {\bf m})'  \vec{\sigma} \; ] =
(a_{0}' -i  a_{0}'  {\bf n}'   \vec{\sigma})  \; ( b_{0}' + b_{0}'
\; { ({\bf m}' +  {\bf n}' \times {\bf m}' ) \over  1 + n^{'2}} \;
\vec{\sigma}   ) \; , \nonumber
\\
 a_{0}' =   \pm    {1 \over  \sqrt{1 +  \lambda^{2} \; n^{2}}  } \; ,
\;\;
  b_{0}' =  {1 \over  \sqrt{1 +  \lambda^{2} \; n^{2}}  }\;,
\;\;
 {\bf n}' \times {\bf m}' =
\lambda^{2}  \; {\bf n} \times {\bf m} \; . \nonumber
\\
\label{44}
\end{eqnarray}

\section{ Behavior of the non-linear constitutive relations\\
under the Lorentz group}

As noted above, in the frame of field theory in non-commutative
space-time,
  extended electrodynamic  equations  minimally modified by  the
first order terms of non-commutativity $\theta_{ab}$ were
constructed  -- those are usual Maxwell equations with  special
non-linear constitutive equations)
\begin{eqnarray}
{ {\bf D} \over \epsilon_{0}} = {\bf E} + [ \; ( {\bf n} {\bf E})
-  ({\bf m} c{\bf B})\;  ]\; {\bf E} +
 [ \; ( {\bf m} {\bf E}) +  ({\bf n} c{\bf B})\;  ]\; c{\bf B} +
 ({\bf E} c{\bf B})\;
  {\bf m}  + {1\over 2}({\bf E}^{2}-
 c^{2}{\bf B}^{2}) \; {\bf n} \; ,
\nonumber
\\
{ {\bf H} \over c \epsilon_{0}} = c {\bf B} + [ \; ( {\bf n} {\bf
E}) -  ({\bf m} c{\bf B})\;  ]\; c{\bf B} -
 [ \; ( {\bf m} {\bf E}) +  ({\bf n} c{\bf B})\;  ]\; {\bf E} -
 ({\bf E} c{\bf B})\;
  {\bf n}  + {1\over 2}({\bf E}^{2}-
 c^{2}{\bf B}^{2}) \; {\bf m} \; ,
\nonumber
\\
\label{46a}
\end{eqnarray}

\noindent and inverse relations
\begin{eqnarray}
 {\bf E} = {\bf D} /\epsilon_{0} +
  [  {\bf m} {\bf H} /c \epsilon_{0} - {\bf n} {\bf D}/ \epsilon_{0}; ]  {\bf D}/ \epsilon_{0} -
[  {\bf m} {\bf D} / \epsilon_{0} + {\bf n} {\bf H} /
c\epsilon_{0}   ]  {\bf H} / c \epsilon_{0} - \nonumber
\\
- ({\bf D} {\bf H} c/\epsilon^{2} _{0})  {\bf m}  +{1 \over 2}
 ( {\bf H}^{2} / c ^{2}\epsilon^{2}_{0} - {\bf D}^{2} /
\epsilon^{2}_{0} ) \; {\bf n} \; ,  \nonumber
\\[3mm]
 c{\bf B} = {\bf H} /c\epsilon_{0} +
  [  {\bf m} {\bf H} /c \epsilon_{0} - {\bf n} {\bf D}/ \epsilon_{0}  ]  {\bf H}/ c\epsilon_{0} +
[  {\bf m} {\bf D} / \epsilon_{0} + {\bf n} {\bf H} /
c\epsilon_{0}   ]  {\bf D} /  \epsilon_{0}  \nonumber
\\
+ ({\bf D} {\bf H} c/\epsilon^{2} _{0})  {\bf n}  +{1 \over 2}
 ( {\bf H}^{2} / c ^{2}\epsilon^{2}_{0} - {\bf D}^{2} /
\epsilon^{2}_{0} )  {\bf m} \; . \nonumber
\\
\label{46b}
\end{eqnarray}

\noindent  In the used system  SI, the dimensions of the
quantities involved obey  relations:
\begin{eqnarray}
[ E ] = [ {D \over \epsilon_{0}} ] = [ cB ] = { H \over c
\epsilon_{0}} ] = [ {1 \over n } ] = [ {1 \over m } ]\; .
\nonumber
\end{eqnarray}

Let us translate these formulas to
 Riemann-Silberstein-Majorana-Oppenheimer  basis (more details and references see in
 \cite{Bogush-Red'kov-Tokarevskaya-Spix-2008}).
Correspon\-ding\-ly, in  the variables with simple transformation
properties under the complex orthogonal group SO(3.C), isomorphic
to the Lorentz  group  $L_{+}^{\uparrow}$. To this end, it
suffices to use the following variables
\begin{eqnarray}
{\bf f} = {\bf E} + ic{\bf B}\;, \qquad {\bf h} = {1 \over
\epsilon_{0}} ({\bf D} + i {\bf H} /c ) \;  , \qquad {\bf n} + i
{\bf m} = {\bf K} \; ; \label{47}
\end{eqnarray}

\noindent the constitutive equations read
\begin{eqnarray}
 {\bf h} = [\; 1 +   (  {\bf f}^{*}  {\bf K}^{*} )  \; ]\; {\bf f} + { ({\bf f}^{*} {\bf f}^{*}) \over 2} \; {\bf K}
\; , \;\;\;  {\bf f} = [\; 1  -  ({\bf h}^{*}  {\bf K}^{*}  )\;]\;
{\bf h} - { {\bf h}^{*} {\bf h}^{*} \over 2 } \; {\bf K} \; .
\label{48a}
\end{eqnarray}

\noindent These relations are inverse to each other within the
accuracy of the first order terms in ${\bf K}$. With respect to
the Lorentz  group the constitutive equations behave themselves as
follows:
\begin{eqnarray}
{\bf h}' = O {\bf h}\; , \qquad {\bf f}' = O {\bf f}\; , \qquad
{\bf f}^{'*} = O^{*} {\bf f}^{*} \; , \qquad  {\bf K}' = O {\bf
K}\; , \qquad {\bf K}^{'*} = O^{*} {\bf K}^{*}  \; , \nonumber
\\
 {\bf h}'  = O [ 1 +
( (O^{*})^{-1} {\bf f}^{'*}  (O^{*})^{-1}{\bf K}^{'*} )   ]
O^{-1}{\bf f}' + O  { ( O^{*})^{-1}{\bf f}^{'*}  (O^{*})^{-1} {\bf
f}^{'*}) \over 2}  O^{-1} {\bf K}'  \; , \nonumber
\\
 {\bf f}'  = O [ 1 -
( (O^{*})^{-1} {\bf h}^{'*}  (O^{*})^{-1}{\bf K}^{'*} )   ]
O^{-1}{\bf h}' - O  { ( O^{*})^{-1}{\bf h}^{'*}  (O^{*})^{-1} {\bf
h}^{'*}) \over 2} O^{-1} {\bf K}' \; . \nonumber
\end{eqnarray}

\noindent Allowing for orthogonality property of the elements of
$SO(3.C)$ we arrive at
\begin{eqnarray}
 {\bf h}'  =  [\; 1 +
( {\bf f}^{'*}  {\bf K}^{'*} )  \; ]\;  {\bf f}' + { {\bf f}^{'*}
{\bf f}^{'*} \over 2} \;  {\bf K}'  , \qquad
 {\bf f}'  =  [\; 1 -
( {\bf h}^{'*}  {\bf K}^{'*} )  \; ]\;  {\bf h}' - { {\bf h}^{'*}
{\bf h}^{'*} \over 2} \;  {\bf K}' \; .
\label{50}
\end{eqnarray}

This means that the constitutive relations are explicitly
covariant  under the complex orthogonal group SO(3.C). Evidently,
above  described 2-parametric small subgroups  in  $SO(3.C)$
leaving invariant non-commutativity parameters, complex 3-vectors
${\bf K}$ and ${\bf K}^{*}$, provide us with subgroup in the
Lorentz group,  leaving invariant the nonlinear constitutive
equations:

\vspace{2mm}

\underline{non-isotropic case}
\begin{eqnarray}
{\bf K} = {\bf n} + i{\bf m} = K \ {\bf k}\; , \qquad {\bf
\Delta}^{2} = 1 \; , \qquad {\bf \Delta} =  {\bf N} + i{\bf M} \; ,
\nonumber
\\
{\bf K}' = O (\gamma , {\bf \Delta}) \; {\bf K} = {\bf K} \; ,
\qquad {\bf K}^{'*} = O^{*} (\gamma , {\bf \Delta}) \; {\bf K}^{*}
= {\bf K}^{*}   \; , \nonumber
\\
\mbox{subgroup} \;\;O (\gamma , {\bf \Delta})\; , \hspace{20mm}
\qquad \gamma'' = \gamma ' + \gamma   \; , \nonumber
\\
 {\bf h}'  =  [\; 1 +
( {\bf f}^{'*}  {\bf K}^{*} )  \; ]\;  {\bf f}' + { {\bf f}^{'*}
{\bf f}^{'*} \over 2} \;  {\bf K}  \; ,\nonumber
\\
 {\bf f}'  =  [\; 1 -
( {\bf h}^{'*}  {\bf K}^{*} )  \; ]\;  {\bf h}' - { {\bf h}^{'*}
{\bf h}^{'*} \over 2} \;  {\bf K} \; , \label{51}
\end{eqnarray}

\underline{isotropic case}
\begin{eqnarray}
{\bf K} = {\bf n} + i{\bf m} =   {\bf \Delta}\; , \qquad {\bf
\Delta}^{2} = 0  \; , \nonumber
\\
{\bf K}' = O ( z {\bf \Delta}) \; {\bf K} = {\bf K} \; , \qquad
{\bf K}^{'*} = O^{*} (\gamma , {\bf \Delta}) \; {\bf K}^{*} = {\bf
K}^{*}  \; ,  \nonumber
\\
\mbox{subgroup} \;\;O ( z\; {\bf \Delta})\; , \hspace{20mm} \qquad
z'' = z' + z   \; , \nonumber
\\
 {\bf h}'  =  [\; 1 +
( {\bf f}^{'*}  {\bf K}^{*} )  \; ]\;  {\bf f}' + { {\bf f}^{'*}
{\bf f}^{*} \over 2} \;  {\bf K}' \; , \nonumber
\\
 {\bf f}'  =  [\; 1 -
( {\bf h}^{'*}  {\bf K}^{*} )  \; ]\;  {\bf h}' - { {\bf h}^{'*}
{\bf h}^{'*} \over 2} \;  {\bf K} \; . \label{52}
\end{eqnarray}

\section{On constitutive relations and discrete dual symmetry
}

In absence of sources, Maxwell equations in media
\begin{eqnarray}
\mbox{div} \; {\bf B} = 0 \; , \qquad \mbox{rot} \;{\bf E} =
-{\partial c{\bf B} \over \partial c  t}  \; ,
\nonumber
\\
\mbox{div}\; {\bf D} = 0 \; , \qquad
 \mbox{rot} \; { {\bf H} \over  c}  =
  {\partial {\bf D} \over \partial c t}
\label{53}
\end{eqnarray}

\noindent  can be combined into complex ones
\begin{eqnarray}
\hspace{30mm} \mbox{div}\; ( {{\bf D}\over \epsilon_{0}}  + i \;c
 {\bf B}) = 0 \; ,
 \nonumber
 \\
 - i \partial_{0}  ( {{\bf D}\over \epsilon_{0}}  + i c {\bf B})
+\mbox{rot}\; ( {\bf E} + i{{\bf H}/c \over \epsilon_{0}} ) = 0 \;
.
 \label{54}
\end{eqnarray}

\noindent  Variables with simple transformation properties under
$SO(3.C)$ are
$$
{\bf f} = {\bf E} + i c {\bf B}   \; , \;  {\bf h} =  {1 \over
\epsilon_{0}} \; ({\bf D} + i {\bf H}  / c )\;  .
$$
Eqs.  (\ref{54}) may be  translated into
\begin{eqnarray}
 \mbox{div} \; (  { {\bf h } + {\bf h}^{*} \over 2 } + { {\bf f} - {\bf f}^{*} \over 2}) = 0\; ,
 \nonumber
 \\
 -i\partial_{0} ( { {\bf h } + {\bf h}^{*} \over 2 } + { {\bf f} -
{\bf f}^{*} \over 2}) + \mbox{rot} \; ({ {\bf f } + {\bf f}^{*}
\over 2 } + { {\bf h} - {\bf h}^{*} \over 2} )  = 0 \; .
\label{57}
\end{eqnarray}

\noindent It has sense to introduce new variables
\begin{eqnarray}
{\bf G} =   { {\bf h } + {\bf f}  \over 2 } \; , \qquad {\bf R} =
{ {\bf h }^{*} - {\bf f}^{*} \over 2 }  \; , \label{58a}
\end{eqnarray}

\noindent they are vectors of different type under $SO(3.C)$
group:
$
{\bf G}' = O \;{\bf G} \; , \;  {\bf R}' = O^{*} \;{\bf R} \;
. $
 Accordingly, Maxwell equations read
\begin{eqnarray}
\mbox{div}\; {\bf G} +  \mbox{div}\; {\bf R} = 0  \; ,
\nonumber
\\
-i\partial_{0} {\bf G}  + \mbox{rot} \; {\bf G} -i\partial_{0}
{\bf R}  - \mbox{rot} \;{\bf R} =  0\; , \label{59}
\end{eqnarray}

\noindent these are invariant under dual rotations:
$
e^{i \chi} \; {\bf G} = {\bf G}' \; , \; e^{i\chi}  \; {\bf R}
= {\bf R}',$
 which can be  translated to variables
 ${\bf h} ,{\bf f}$:
\begin{eqnarray}
{\bf h}' =\cos \chi \; {\bf h} + i \; \sin \chi \; {\bf f} \;  ,
\qquad {\bf f}' = i\; \sin  \chi \; {\bf h} + \; \cos \chi \; {\bf
f} \; . \nonumber
\end{eqnarray}

\noindent Following to \cite{Aschiery-2001}, the dual rotations
for ${\bf K}$ is taken in the form
$
{\bf K} ' = e^{i \chi} \; {\bf K} \; .
$
Let us consider behavior of the constitutive relations with
respect to the dual rotation (for brevity, let $\epsilon_{0} = 1, c =
1$):
\begin{eqnarray}
 {\bf h} =
[\; 1 +   ({\bf f}^{*} {\bf K}^{*} )  \; ]\; {\bf f} + { ({\bf
f}^{*} {\bf f}^{*}) \over 2} \; {\bf K}\; , \qquad
 {\bf f}  =  [\; 1 -
( {\bf h}^{*}  {\bf K}^{*} )  \; ]\;  {\bf h} - { {\bf h}^{*} {\bf
h}^{*} \over 2} \;  {\bf K}\; .
 \nonumber
\end{eqnarray}

\noindent  We immediately note three discrete operations leaving
invariant the constitutive
 relations:

 \begin{eqnarray}
\underline{(1) \qquad \chi = {\pi \over 2}\; ,}
 \qquad {\bf h}' = i \; {\bf f}\;, \;\; {\bf f}' = i \; {\bf h}\;, \;\;
{\bf K} ' = i \; {\bf K}  \; , \nonumber
\\
 {\bf f}'  =  [\; 1 -
( {\bf h}^{'*}  {\bf K}^{'*} )  \; ]\;  {\bf h}' - { {\bf h}^{'*}
{\bf h}^{'*} \over 2} \;  {\bf K}'  \; , \qquad
 {\bf h}' =
[\; 1 +   ({\bf f}^{'*} {\bf K}^{'*} )  \; ]\; {\bf f} '+ { ({\bf
f}^{'*} {\bf f}^{'*}) \over 2} \; {\bf K}' \; , \nonumber
\\[5mm]
\underline{(2) \qquad \chi = \pi \; ,}
 \qquad {\bf f}' = - \; {\bf f}\;, \;\; {\bf h}' = - \; {\bf h}\;, \;\;
{\bf K} ' = - \; {\bf K} \; , \nonumber
\\
 {\bf h}' =
[\; 1 +   ({\bf f}^{'*} {\bf K}^{'*} )  \; ]\; {\bf f}' + { ({\bf
f}^{'*} {\bf f}^{'*}) \over 2} \; {\bf K}' \; , \qquad
 {\bf f}'  =  [\; 1 -
( {\bf h}^{'*}  {\bf K}^{'*} )  \; ]\;  {\bf h}' - { {\bf h}^{'*}
{\bf h}^{'*} \over 2} \;  {\bf K}' \; , \nonumber
\\[5mm]
\underline{(3) \qquad \chi = {3\pi \over 2}\; ,}
 \qquad {\bf h}' = -i \; {\bf f}\;, \;\; {\bf f}' = -i \; {\bf h}\;, \;\;
{\bf K} ' = -i \; {\bf K}  \; , \qquad
\nonumber
\\ {\bf f}'  =  [\; 1 -
( {\bf h}^{'*}  {\bf K}^{'*} )  \; ]\;  {\bf h}' - { {\bf h}^{'*}
{\bf h}^{'*} \over 2} \;  {\bf K}' \; ,
 {\bf h}' =
[\; 1 +   ({\bf f}^{'*} {\bf K}^{'*} )  \; ]\; {\bf f} '+ { ({\bf
f}^{'*} {\bf f}^{'*}) \over 2} \; {\bf K}' \; .
\nonumber
\\
\label{64}
\end{eqnarray}

\noindent Together with the unit transform
\begin{eqnarray}
\underline{(4) \qquad \chi = 0 \; ,}
 \qquad {\bf f}' = + \; {\bf f}\;, \;\; {\bf h}' = + \; {\bf h}\;, \;\;
{\bf K} ' = + \; {\bf K}  \nonumber
\end{eqnarray}

\noindent  we have  the discrete group of four element with simple
structure: $\{\; 1 , -1, +i , \; - i \; \} \; . $

Let us consider action of continuous dual rotations on
constitutive equations. It is convenient to use the variables
 ${\bf G}, {\bf R}$:
\begin{eqnarray}
{\bf G} + {\bf R}^{*} =  {\bf h}\; , \qquad {\bf G}^{*} + {\bf R}
=  {\bf h}^{*} \; ,
\nonumber
\\
{\bf G} - {\bf R}^{*} =  {\bf f}\; , \
\nonumber
\\
 {\bf G}^{*} - {\bf R}
=  {\bf f}^{*}\; , \label{66}
\end{eqnarray}

\noindent then
\begin{eqnarray}
{\bf G} + {\bf R}^{*} = [\; 1 +   ({\bf G}^{*} - {\bf R})  {\bf
K}^{*} )  \; ]\;  ({\bf G} - {\bf R}^{*})  +
 { ({\bf G}^{*} - {\bf R})({\bf G}^{*} - {\bf R})   \over 2} \; {\bf K} \; ,
\nonumber
\\
{\bf G} - {\bf R}^{*}   =  [\; 1 - ( {\bf G}^{*} + {\bf R} )  {\bf
K}^{*} )  \; ]\;  ( {\bf G} + {\bf R}^{*})  - {  ({\bf G}^{*} +
{\bf R} ) ({\bf G}^{*} + {\bf R} )  \over 2} \;  {\bf K} \; ,
\nonumber
\end{eqnarray}

\noindent from whence it follows
\begin{eqnarray}
 2 ( {\bf G}^{*} {\bf R} )\; {\bf K}  +  ({\bf G}^{*} {\bf K}^{*}) \; {\bf R}^{*} +
  ({\bf R} {\bf K}^{*}) \; {\bf G} =0 \; ,
\nonumber
\\
2{\bf R}^{*} =  ({\bf G}^{*} {\bf K}^{*})\;  {\bf G}  + ({\bf R}
{\bf K}^{*})\; {\bf R}^{*} + {1 \over 2} \; (  \; {\bf G}^{*} {\bf
G}^{*}  + {\bf R} {\bf R} \; ) \; {\bf K} \; . \label{68}
\end{eqnarray}

\noindent When  ${\bf K}=0$, eq.   (\ref{68}) gives $ 0 =0 \; , \;
{\bf R}= 0 , \; \Longrightarrow \; {\bf h}= {\bf f} , $
 which coincides with  constitutive relations in vacuum.
With respect to the dual rotation
\begin{eqnarray}
e^{i \chi} \; {\bf G} = {\bf G}'  \; , \;  e^{-i \chi} \; {\bf
G}^{*} = {\bf G}^{'*} \;, \;\;
e^{i \chi} \; {\bf R} = {\bf R}'  \; ,
\nonumber
\\
  e^{-i \chi} \; {\bf
R}^{*} = {\bf R}^{'*} \;,
\;\;
e^{i \chi} \; {\bf K} = {\bf K}'  \; , \;  e^{-i \chi} \; {\bf
K}^{*} = {\bf K}^{'*} \nonumber
\end{eqnarray}

\noindent eqs.  (\ref{68}) transform into
\begin{eqnarray}
 2 \; ( e^{i \chi}  {\bf G}^{'*} \; e^{-i\chi} {\bf R}' )\; e^{-i\chi} {\bf K}' \;
 \nonumber
 \\
 + \;
 ( e^{i \chi}  {\bf G}^{'*} \;   e^{i\chi} {\bf K}^{'*}) \; e^{i\chi}  {\bf R}^{'*} \;  +
  ( e^{-i\chi} {\bf R}'   \;  e^{i\chi} {\bf K}^{'*}  ) \; e^{-i\chi} {\bf G}' = 0 \; ,
\nonumber
\\[5mm]
2 \; e^{i\chi}  {\bf R}^{'*}  =  ( e^{i \chi}  {\bf G}^{'*}  \;
e^{i\chi} {\bf K}^{'*})\; e^{-i\chi} {\bf G}'  \; + ( e^{-i\chi}
{\bf R}'  \; e^{i\chi} {\bf K}^{'*}  )\;  e^{i\chi}  {\bf
R}^{'*} \nonumber
\\
+ {1 \over 2} \; [  \; e^{i \chi}  {\bf G}^{'*} \; e^{i \chi} {\bf
G}^{'*}  + e^{-i\chi} {\bf R}'\; e^{-i\chi} {\bf R}'  \; ] \;
e^{-i\chi} {\bf K}' \; . \nonumber
\end{eqnarray}

\noindent  Requiring invariance of these equations we arrive at
two simple  equations with evident solution
\begin{eqnarray}
e^{-i\chi} = e^{+3i\chi}\; , \;  e^{+i\chi} = e^{-3i\chi}\;, \;
e^{+4i\chi} = 1\; ,
\nonumber
\\
 e^{i\chi} = 1,-1, +i, -i \; .
\label{70}
\end{eqnarray}

Therefore, only discrete dual transformation leaves invariant the
non-linear constitutive  equations,
 it corresponds to  $e^{i\chi}= \pm i $.
 Thus, the dual symmetry status in non-commutative electrodynamics
differs  with that in  ordinary   linear Maxwell theory in
commutative space, this fact is  to be  interpreted in physical
terms.

\section{Acknowledgement}

Authors are grateful to Professor Kurochkin Ya.A. for discussion
and advice. This  work was    supported  by the Fund for Basic
Research of Belarus, grant F08R-039.

\end{document}